%% file: current_cumulants_2.tex
\documentclass[aps,pre,10pt,twocolumn,twoside,a4paper,floatfix,superscriptaddress,showkeys,showpacs]{revtex4-1}

\usepackage{graphicx}
\usepackage{amsmath,amssymb,amsfonts}
\usepackage{mathtools}
\usepackage{bm}

\usepackage[amsmath,hyperref,thmmarks]{ntheorem}

\usepackage{siunitx}
\usepackage{booktabs} 

\usepackage{time}
\usepackage{enumerate}

\input{current_cumulants_macros}

\begin{document}

\date{\today{}}

\title{Fluctuating Currents in Stochastic Thermodynamics II.\\
  Energy Conversion and Nonequilibrium Response in Kinesin Models} 

\author{Bernhard Altaner}
\affiliation{Max Planck Institute for Dynamics and Self-Organization (MPI DS), Am Fassberg 17, 37077 G\"{o}ttingen, Germany}
\affiliation{Institute for Nonlinear Dynamics, Faculty of Physics, Georg-August University G\"{o}ttingen, 37077 G\"{o}ttingen, Germany}
\author{Artur Wachtel}
\affiliation{Max Planck Institute for Dynamics and Self-Organization (MPI DS), Am Fassberg 17, 37077 G\"{o}ttingen, Germany}
\affiliation{Complex Systems and Statistical Mechanics, Physics and Materials Science Research Unit, University of Luxembourg, Luxembourg}
\author{J\"{u}rgen Vollmer}
\affiliation{Max Planck Institute for Dynamics and Self-Organization (MPI DS), Am Fassberg 17, 37077 G\"{o}ttingen, Germany}
\affiliation{Institute for Nonlinear Dynamics, Faculty of Physics, Georg-August University G\"{o}ttingen, 37077 G\"{o}ttingen, Germany}

\begin{abstract}
 Unlike macroscopic engines, the molecular machinery of living cells is strongly affected by fluctuations.
 Stochastic Thermodynamics uses Markovian jump processes to model the random transitions between the chemical and configurational states of these biological macromolecules.
 A recently developed theoretical framework [Wachtel, Vollmer, Altaner: ``Fluctuating Currents in Stochastic Thermodynamics I. Gauge Invariance of Asymptotic Statistics''] provides a simple algorithm for the determination of macroscopic currents and correlation integrals of arbitrary fluctuating currents.
 Here, we use it to discuss energy conversion and nonequilibrium response in different models for the molecular motor kinesin.
 Methodologically, our results demonstrate the effectiveness of the algorithm in dealing with parameter-dependent stochastic models.
 For the concrete biophysical problem our results reveal two interesting features in experimentally accessible parameter regions:
 The validity of a non-equilibrium Green--Kubo relation at mechanical stalling as well as a negative differential mobility for superstalling forces.
\end{abstract}

\maketitle

\section{Introduction}
\label{sec:intro}

Understanding the complex biochemical processes which are responsible for cellular metabolism is one of the key questions in modern biophysics.
The quantitative analysis of so-called molecular motors, which are the small machines transforming different forms of energy into one another, is at the center of these efforts~\cite{Ritort2004,Bustamante.etal2005,Seifert2008}.
In recent years scientists developed techniques that allow the systematic observation and manipulation of these biological macromolecules~\cite{Ritort2006}.
Under in vivo conditions, \text{(electro-)}chemical gradients in the cell maintain these systems out of equilibrium.
From a thermodynamic perspective, one is interested in the currents of heat, matter and energy that flow through a molecular motor, because they allow, for instance, the definition of its efficiency.

In analogy to macroscopic engines, molecular motors are described by thermodynamic cycles in a space of biochemical and configurational states.
In contrast, the energy scales involved in biochemical energy conversion are only a couple of times larger than the thermal energy. 
Consequently, thermal fluctuations cannot be neglected, and their dynamics must be modeled as a stochastic process that reproduces the stochastic time series observed in experiments.
Stochastic Thermodynamics refers to a general framework for a consistent definition of fluctuating work and heat currents on the level of these fluctuating time series~\cite{Sekimoto1998,Seifert2012}.
The common model for molecular motors are dynamically reversible Markov jump processes, which can be thought of as (memoryless) random walks on a biochemical network of states \cite{Hill1966,Schnakenberg1976,Seifert2008,Esposito.VandenBroeck2010}.
In an accompanying publication~\cite{WachtelVollmerAltaner2015} we investigated the asymptotic statistics of such systems from the perspective of the cycle topology of the network of states.
In particular, we developed an efficient method to calculate all cumulants of arbitrary fluctuating currents analytically.

Here, we are interested in the first and second order fluctuation statistics, \ie{} the expressions for macroscopic average currents (like the motor's velocity) and Green--Kubo time-correlation integrals (like its diffusion constant).
To be concrete, we use the analytic nature of our method to analyze the parameter space of different stochastic models for the motor protein kinesin~\cite{Liepelt+Lipowsky2007,Lau.etal2007,Altaner.Vollmer2012}, which were designed to reflect typical force-spectroscopy experiments~\cite{Schnitzer.Block1997,Visscher.etal1999,Carter.Cross2005,Clancy.etal2011}.
Besides illustrating the insights that thermodynamic cycles provide into the motor dynamics, our results uncover interesting model predictions and thus indicate directions for future experimental research:
The validity of a nonequilibrium fluctuation dissipation relation at mechanical stalling as well as negative differential mobility, commonly referred to as ``getting more from pushing less'' \cite{Zia2007}.

This work is structured as follows.
In Sec.~\ref{sec:basics} we briefly review the results of Ref.~\cite{WachtelVollmerAltaner2015}.
In contrast to the formal exposition there, here we focus on the implementation of a universally applicable algorithm for the efficient calculation of averages and correlation integrals of fluctuating currents in Stochastic Thermodynamics.
Sec.~\ref{sec:kinesin} thoroughly discusses how to apply our universal method in the concrete biophysical context of a kinesin model.
In Sec.~\ref{sec:results} we give a detailed account of kinesin's chemical (ATP hydrolysis) and mechanical (displacement) currents as functions of their conjugate chemical and mechanical drivings.
We conclude in Sec.~\ref{sec:conclusion} with a discussion of the main conceptional and biophysical insights.

\section{Fluctuating currents and their statistics}
\label{sec:basics}
In this section we introduce our mathematical notation and -- based on the general results presented in Ref.~\cite{WachtelVollmerAltaner2015} -- provide a concise recipe for calculating the averages and asymptotic (co-)variances of two fluctuating currents in a dynamically-reversible Markov process on a finite state space.
Such averages and covariances play a major role in Stochastic Thermodynamics  \cite{Hill1977,Schnakenberg1976,Maes2004}:
They correspond to physical steady-state currents and time-correlation (Green--Kubo) integrals \cite{Lebowitz+Spohn1999}.
To be concrete, we exemplify topological concepts for both a four-state and a six-state Markov process, \Fig{trees-cycles}.
In \Sec{kinesin} we interpret these examples as models for the molecular motor kinesin.

\subsection{Currents for Markovian processes}
\label{sec:jump-observables}
Memoryless stochastic processes on a finite state space \(\vertices=\left\{ v_1, v_2, \dots, v_N \right\}\) are called \emph{Markovian jump processes}.
Henceforth, we consider the time-continuous and homogeneous case.
A realization, or \emph{trajectory} $(\tray_k,t_k)$,  of the process starting at time \(t_0=0\) in a state \(\tray_0\in\vertices\) is a collection of jump times \(t_k>0\) and visited states \(\tray_k\in\vertices\) with \(k\in\naturals\).
We interpret it as a time series that contains the outcomes of subsequent measurements performed on a small system like a molecular motor.

\begin{figure}
  \centering
  \includegraphics[width=.48\textwidth]{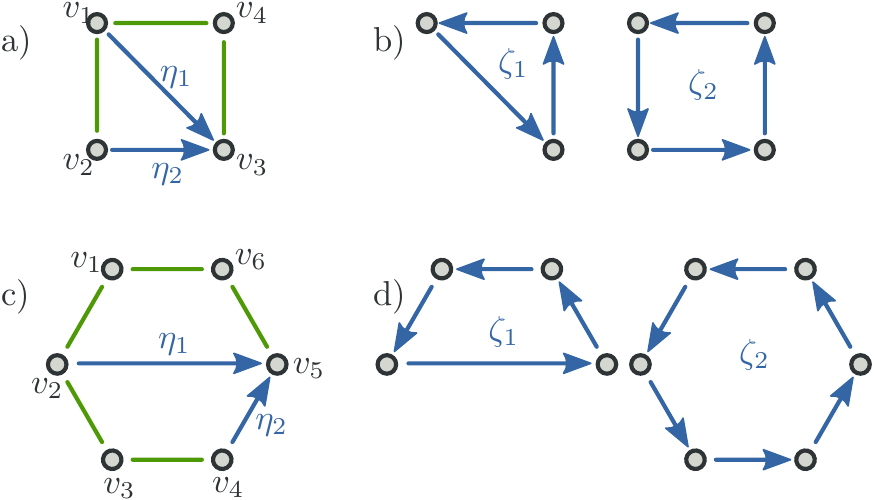}
  \caption{
    Two different graphs representing Markov models with (a) four states ($N=4$, $M=5$) and (c) six ($N=6$, $M=7$) states.
    The edges marked in green serve as a spanning tree \(\tredges\).
    They connect all vertices of the respective graph.
    The remaining edges, marked blue, are the respective chords \(\chords\).
    Here, we already indicate an orientation for the chords to provide a reference for the sign of the currents.
    Each chord \(\eta_\ell\) gives rise to a fundamental cycle \(\zeta_\ell\).
    They are shown in panels (b)  and (d) for the four and six-state models, respectively.
    Regarding the topological cycle structure, both graphs are equivalent.
    In particular, they have the same number \(B=M-N+1=2\) of fundamental cycles.
}
  \label{fig:trees-cycles}
\end{figure}

Transitions from a state \( v_i \in\vertices \) to a different state \( v_j\in\vertices\) occur at a given constant rate \(w^i_j\).
For thermodynamic consistency~\cite{Hill1966,Schnakenberg1976,Lebowitz+Spohn1999,Maes2004,Seifert2012}, we require \emph{dynamical reversibility}, \ie{} 
\( w^i_j > 0 \Leftrightarrow w^j_i > 0 \)\,.
With this constraint we draw the state space as an undirected graph \(\graph\) with the \(N\) states as vertices and \(M\) admissible transitions as edges, \cf{} \Fig{trees-cycles}.
In addition to dynamic reversibility we assume that the state space is connected, which ensures ergodicity of the process.

An ensemble of trajectories with initial probability distribution \(\vec{p}(0)=\left( p_1(0), \dots, p_N(0) \right)\) on \(\vertices\) evolves according to the master equation~\citep{vKampen1992}: \( \frac{\D}{\D t} \vec{p}(t) = \vec{p}(t)\tmat \) or, in components,
\[ \frac{\D}{\D t} p_i(t) = \sum_{j\neq i} \left[ p_j w^j_i - p_i w^i_j \right] \,,\]
where we use the convention \(w^i_i = - \sum_{j\neq i} w^i_j \)\,.
Ergodicity of the process implies that there is a unique steady-state probability distribution \(\vec{\pi}\) satisfying \(0 = \vec{\pi}\tmat \) 
 to which all initial conditions will converge eventually.
The quantities \(J^i_j = \pi_i w^i_j - \pi_j w^j_i\) represent the steady-state probability currents between two states \(v_i\) and \(v_j\).
The special condition where all the steady-state currents vanish, \(J^i_j = 0\), is called \emph{detailed balance} or \emph{equilibrium}.
Since we are interested in the currents of non-equilibrium systems, we will \emph{not} assume detailed balance in the following.

In order to account for general currents, \eg{} changes in energy, entropy, particle numbers or physical position, we introduce \emph{jump observables}.
A jump observable \(\obs\) assigns a weight \(\obs^i_j\) to the transition from \(v_i\) to \(v_j\), where we require anti-symmetry: \( \obs^i_j = - \obs^j_i \).
The \emph{macroscopic average current} $\ssav{\obs}$ associated to a jump observable \(\obs\) is
\begin{align}
  \ssav{\obs} \coloneqq \frac{1}{2} \sum_{i,j} J^i_j\, \obs^i_j\,. \label{eq:macro-current}
\end{align}

In order to illustrate the concept we provide two examples.
For a pair of states \( \left( v_i, v_j \right)\) we define a simple but important case of jump observable: 
the \emph{counting observable} \(\bobs{i}{j}\) counts \(+1\) for a transition from state \(v_i\) to \(v_j\) and \(-1\) for a reverse transition from \(v_j\) to \(v_i\).
To every other transition it associates a weight of \(0\). 
In this case the macroscopic counting rate from \(v_i\) to \(v_j\) equals the steady-state probability current between these states: \( \ssav{\bobs{i}{j}} = J^i_j \)\,.
This expression obviously vanishes if transitions between \(v_i\) and \(v_j\) are impossible.
In Ref.~\citep{WachtelVollmerAltaner2015} we emphasized that counting observables form a basis of the space of jump observables:
every jump observable can be expressed as an appropriate linear combination of counting observables.

Another important example is the dissipation in stochastic thermodynamics.
It is derived from the jump observable \(\sigma\) that takes the values \(\sigma^i_j = \ln\frac{w^i_j}{w^j_i}\).
The macroscopic average dissipation  
\[ \ssav{\sigma} = \frac{1}{2} \sum_{i,j} J^i_j \ln\frac{w^i_j}{w^j_i}\]
is non-negative and vanishes only at equilibrium, \ie{} if and only if we have detailed balance. 

Kirchhoff's current law states that the currents in an electrical network balance at each vertex $v_i$.
The same is true for the steady-state probability currents $J^i_j$, and the stationary Master equation formalizes Kirchhoff's \emph{current law} as \(\sum_i J^i_j =0 \)\,.
In Ref.~\citep{Schnakenberg1976}, Schnakenberg discussed an extended analogy between Markov jump processes and Kirchhoff's laws. 
The fundamental object of Schnakenberg's theory is a set of \(B=M-N+1\) fundamental cycles \(\{\zeta_{\ell}\}\), which describe the topology of a Markov jump process.
They are obtained from a spanning tree of the graph representing the network of states, \cf{} \Fig{trees-cycles} as well as \Sec{algorithm}.
For now, think of a fundamental cycle \(\zeta_\ell\) as a tuple of consecutive states \( \left( \tray_0, \tray_1, \dots, \tray_{m(\ell)}=\tray_0 \right) \) that form a self-avoiding and closed trajectory.
Cycles are defined up to cyclic permutations.
Adding the contributions of a jump observable $\obs$ along the transitions in a fundamental cycle \( \zeta_\ell\) gives its \emph{circulation}
\begin{align}
	\circu{\obs}{\ell} \coloneqq \sum_{k = 0}^{m(\ell)} \obs^{\tray_k}_{\tray_{k+1}} \label{eq:affinity}\,.
\end{align}
Kirchhoff's \emph{voltage law} states that the voltage drops $V = U_j - U_i$ between vertices of an electric network vanish if integrated along any circuit.
Using the notion of circulations, the voltage law reads $\circu{V}{l}=0$.
Another result obtained in this context is the  Schnakenberg decomposition of the macroscopic dissipation rate~\cite{Schnakenberg1976}:
\begin{align}
	\ssav{\sigma} = \sum_{\ell=1}^{B} c_\ell \,  \circu{\sigma}{\ell},
  \label{eq:schnakenberg-decomposition}
\end{align}
where \(c_{\ell}\) is the steady-state probability current associated to the cycle \(\zeta_\ell\)~\cite{WachtelVollmerAltaner2015}.
The circulations $\circu{\sigma}{\ell}$ of the dissipation are called the \emph{cycle affinities}.
In the context of irreversible thermodynamics~\cite{Groot1984}, the Schnakenberg decomposition \eqref{eq:schnakenberg-decomposition} identifies the cycle affinities as the generalized forces which are conjugate to the cycle currents \(c_{\ell}\). 

In an earlier publication~\citep{Altaner.Vollmer2012} the authors pointed out that Schnakenberg's decomposition is equally applicable to other observables \(\obs\).
As a corollary, one may express the average current \(\ssav{\obs}=\sum_{\ell} c_{\ell}\, \circu{\obs}{\ell}\) using only the cyclic structure of the graph.
\Fig{trees-cycles} shows two graphs with four and six states, respectively, but having the same cyclic structure.
Collecting all \(B\) circulations of an observable \(\obs\) in a $B$-tuple gives its \emph{chord representation} \(\obs_\chords \coloneqq \left( \circu{\obs}{1}, \dots, \circu{\obs}{B} \right) \in \reals^B\).
The detailed mathematical background of this representation is discussed in Ref.~\cite{WachtelVollmerAltaner2015}.

Here we are interested not only in the macroscopic expectations of currents, but also in their higher order statistics.
Consequently, the object of study in this work are \emph{fluctuating currents}.
The \emph{instantaneous current} \(\jmath_{\obs}(t)\) derived from a \emph{jump observable} \( \obs \) along a trajectory \( \left( \tray_k, t_k \right) \) is defined as 
\begin{align*}
  \jmath_{\obs}(t) = \sum_{k=0}^{\infty} \delta(t-t_k)\, \obs^{\tray_{k-1}}_{\tray_{k}} \,.
\end{align*}
The \emph{time-integrated current}
\begin{align}
  \obs(T) \coloneqq \int_{t=0}^{T} \jmath_{\obs}(t) \,\D t
  = \sum_{k=0}^{n(T)} \obs^{\tray_{k-1}}_{\tray_{k}}
\end{align}
thus accounts for the total change of the observable \(\obs\) along a random trajectory with a random number \(n(T)\) of jumps up to time \(T\).
Hence, this time-integral is a random variable with its own statistics.
A typical realization of the time-integrated current, and thus its expectation value, grow linearly in time. 
Due to ergodicity, the \emph{time-averaged current} \(\tav{\obs} \coloneqq \frac{1}{T} \obs(T) \) converges to the macroscopic current $\ssav{\obs}$ in the long time limit:
\begin{align}
  \tav{\obs} \xrightarrow{T\to\infty} \ssav{\obs} \,. \label{eq:tav-convergence}
\end{align}
Equivalently, one can average the fluctuating current over trajectories in the steady-state ensemble: \( \left\langle \jmath_{\obs}(t) \right\rangle = \left\langle \jmath_{\obs}(0) \right\rangle = \ssav{\obs} \)\,, where one exploits the fact that the steady state is time independent. 
This ensures that the macroscopic current \(\ssav{\obs}\) is the expectation value, or mean, of the fluctuating current \(\jmath_{\obs}(t)\).

Another important statistical measure are the correlations of two currents \(\jmath_{\obs}\) and \(\jmath_{\oobs}\).
A measure for this correlation is the \emph{Green--Kubo integral}:
\begin{align}
	\ssav{\obs,\oobs} \coloneqq \int_{t=0}^{\infty}
	\trav{\left(\jmath_{\obs}(0) - \ssav{\obs}\right)\left( \jmath_{\oobs}(t) - \ssav{\oobs} \right)}
	\,\D t\,. \label{eq:correlation}
\end{align}
Similar to the case of the average currents, ergodicity allows us to replace the steady-state ensemble average by a time average over the argument of the first current \(\jmath_{\obs}\).
As a consequence~\cite{Lebowitz+Spohn1999}, the correlation integral corresponds to a properly scaled covariance of the
time-averaged currents:
\begin{align}
  \ssav{\obs,\oobs} 
  &= \lim_{T\to\infty} T \Cov\left[ \tav{\obs},\tav{\oobs} \right]\,. \label{eq:green-kubo}
\end{align}
As such, the macroscopic current and the Green--Kubo integral are the first two \emph{scaled cumulants} of the pair \( \left( \tav{\obs}, \tav{\oobs} \right)\) of time-averaged currents.
Scaled cumulants are defined as derivatives of the \emph{scaled cumulant-generating function}, which can be obtained using methods from Large Deviation theory~\citep{Lebowitz+Spohn1999,Andrieux2007,Touchette2009,WachtelVollmerAltaner2015}.
Higher order derivatives represent higher orders of the statistics, such as skewness and kurtosis.

In our accompanying publication~Ref.~\citep{WachtelVollmerAltaner2015} we prove a gauge invariance of the fluctuation statistics and show in detail how Schnakenberg's decomposition is extended to all cumulants of arbitrary observables.
In the next section we give a brief review in form of a concise and efficient recipe for the first two orders.

\subsection{Determining averages and (co-)variances of currents}
\label{sec:algorithm}
The only ingredients needed for the calculation of the scaled cumulants are the transition matrix \(\tmat\) and the jump observables representing the currents of interest.
The elements $w^i_j$ of the transition matrix as well as the jump observables may depend on the (physical) control parameters of the Markov processes in an arbitrary way.
In the following, we present a simple yet efficient algorithm for the calculation of the first two scaled cumulants of two jump observables \(\obs\) and \(\oobs\).
It consists of three sub-steps: topological, algebraic and physical.
The first two steps are universal. 
Only the last step involves the jump observables in question.
Note that the algorithm does neither require the steady-state distribution \(\vec{\pi}\) nor the scaled-cumulant generating function.
We emphasize this fact, because in general these quantities are difficult or even impossible to obtain analytically, \ie{} in the form of a fully parameter-dependent, symbolic expression.

\subsubsection{Topology: defining fundamental cycles}

The first step in the analysis addresses the topology of the graph \(\graph\) representing a network of states, \cf{} \Fig{trees-cycles}. 
\begin{enumerate}[ a. ]
  \item Choose a spanning tree \(\tredges\) for the undirected graph, \ie{} an undirected subgraph spanning all vertices but not containing any circuit
    (green edges in \Figs{trees-cycles}a and c).
  \item Provide an orientation to the \(B= N-M+1\) undirected edges \(\eta_\ell\in\chords\) left out by the tree \(\tredges\). 
    They are the called \emph{chords} (blue edges in \Figs{trees-cycles}a,c).
  \item Identify the fundamental cycles: for every chord \(\eta_\ell \in \chords\), its terminus and origin are connected by a unique directed path through the spanning tree.
    Adding the chord itself as a closure of this path results in the fundamental cycle \(\zeta_\ell\) 
    (\Figs{trees-cycles}b,d).
\end{enumerate}

\subsubsection{Algebra: determining the fundamental current cumulants}
\label{sec:fundamentalCumulants}

The second step of our algorithm involves the determination of the first two (joint) scaled cumulants of the fluctuating currents associated to the chords \(\eta_\ell \in \chords\).

\begin{enumerate}[ a. ]
  \item Write down the characteristic polynomial \(\chi_\chords(\lambda;q_1,q_2,\dots,q_B)=\det(\tmat_\chords-\lambda\mathbb{I})\) of the matrix \(\tmat_\chords\) with entries
    \begin{align}
      \left( \tmat_\chords \right)^i_j =
      \begin{cases}
	w^i_j \exp\left( \pm q_\ell \right) & \text{if } (i\to j) = \pm \eta_{\ell},\\
        w^i_j & \text{else}.
      \end{cases}
      \label{eq:tilted-current-generator}
    \end{align}
  \item Identify the coefficients \(a_0(\vec q)\), \(a_1(\vec{q})\) and \(a_2(\vec{q})\) of \(\chi_\chords(\lambda;\vec{q})=\sum_{k=0}^{N}a_k(\vec{q})\,\lambda^k\), \ie{} the coefficients of the constant, the linear and the quadratic term.
  \item Calculate the vector \(\vec{c}\in\reals^B\) with entries \(c_\ell=\ssav{\eta_\ell}\) and the scaled co-variance matrix \(\vec{C}\in\reals^{B\times B}\) with entries \(C_{\ell m} = \ssav{\eta_\ell,\eta_m}\), as follows:
\begin{subequations}
  \begin{align}
    c_\ell &= -\frac{\partial_\ell a_0}{a_1}, \label{eq:c1fundamental}\\
    C_{\ell m} &=  -\frac{\partial^2_{\ell m} a_0}{a_1} -\frac{2(\partial_\ell a_0)(\partial_{m} a_0)a_2}{a_1^3}\nonumber\\
    &\phantom{-\frac{\partial^2_{\ell m} a_0}{a_1}} + \frac{(\partial_{m} a_1)(\partial_\ell a_0)+(\partial_\ell a_1)(\partial_{m} a_0)}{a_1^2} \nonumber\\
    &= -\frac{\partial^2_{\ell{m}}a_0 + (\partial_\ell a_1) c_{m} +(\partial_{m} a_1)c_\ell +2 a_2 c_{m} c_\ell}{a_{1}},\label{eq:c2fundamental}
  \end{align}
  \label{eq:cfundamental}%
\end{subequations}%
where the partial derivatives \(\partial_\ell a_k \coloneqq \left.\frac{\partial a_k(\vec q)}{\partial q_\ell}\right\vert_{\vec q = \vec{0}}\) and the coefficients \(a_k\) are evaluated at \(\vec q = \vec0\).
\end{enumerate}

\revi{\textit{Remark:} Higher order scaled cumulants are similarly accessible.
The characteristic equation \(0=\chi_\chords(\lambda;\vec{q})\) uniquely defines the entire scaled cumulant-generating function \(\lambda_\chords(\vec{q})\) with \(\lambda_\chords(\vec{0})=0\).
Taking derivatives of the characteristic equation yields linear equations for the cumulants, \ie{} the inner derivatives \(\del_{q_i, \dots , q_j}\lambda_\chords(\vec{0})\).
Note that higher order cumulants depend on the coefficients \(a_k(q)\) with \(k>2\), and the symbolic expressions become more complex.
The first two orders are explicitly given by Eqs.~\eqref{eq:cfundamental}.
The symbolic manipulations that are necessary to obtain the higher orders are efficiently implemented in modern computer algebra systems.
For more details on the procedure and a derivation of Eqs.~\eqref{eq:cfundamental}, the reader is referred to our accompanying publication~\cite{WachtelVollmerAltaner2015}.
}
%

\subsubsection{Physics: cumulants of jump observables}
\label{sec:appendix-observables}

The third and final step of the algorithm yields the first two scaled cumulants of the fluctuating currents associated to the jump observables $\obs$ and $\oobs$.

\begin{enumerate}[ a. ]
  \item Sum the jump observables \(\obs\) and \(\oobs\) along the edges of the fundamental cycle \(\zeta_\ell\) to obtain the circulations \(\circu{\obs}{\ell}\) and \(\circu{\oobs}{\ell}\).
    They are the coordinates of the chord representations \(\obs_\chords, \oobs_\chords\in\reals^B\).
  \item The steady-state average of \(\obs\), and of the scaled covariance of \(\obs\) and \(\oobs\) then read
    \begin{subequations}
    \begin{align}
	    \ssav{\obs} &= \vec{c} \cdot \obs_\chords \equiv \sum_{\ell = 1}^B \circu{\obs}{\ell}\, c_\ell
      \label{eq:multilinear-magic-1}\\
      \ssav{\obs,\oobs} &= \obs_\chords \cdot \vec{C}\, \oobs_\chords \equiv \sum_{{m},\ell =1}^{B} \circu{\obs}{\ell}\, C_{\ell{m}}\, \circu{\oobs}{m}
      \label{eq:multilinear-magic-2}
    \end{align}
    \label{eq:multilinear-magic}
    \end{subequations}
\end{enumerate}

We conclude the section with final remarks on the choice of the spanning tree in step 1, which is a priori arbitrary.
Different choices yield different chords -- and thus different expressions for the fundamental current vector $\vec{c}$ and the fundamental covariance matrix $\vec{C}$.
Expressions \eq{cfundamental} and \eq{multilinear-magic} are universal.
In order to calculate the cumulants of any jump observable, no equations need to be solved.
Via \Eq{cfundamental}, any combinatorial complexity is hidden in (the derivatives of) the coefficients $a_k$ of the characteristic polynomial.
The latter are calculated in a straightforward way either manually or by using a computer-algebra system.
However, the final expressions \eq{multilinear-magic} have fewer terms if some of the circulations \(\circu{\obs}{\ell}\) or \(\circu{\oobs}{\ell}\) along fundamental cycles vanish.
It is thus worthwhile to take a careful look at the particular set of jump observables $\obs$ and $\oobs$ under consideration and choose a spanning tree that is optimal in that regard.


\section{Kinesin}
\label{sec:kinesin}
\begin{figure}[tbp]
 \includegraphics[width=.48\textwidth]{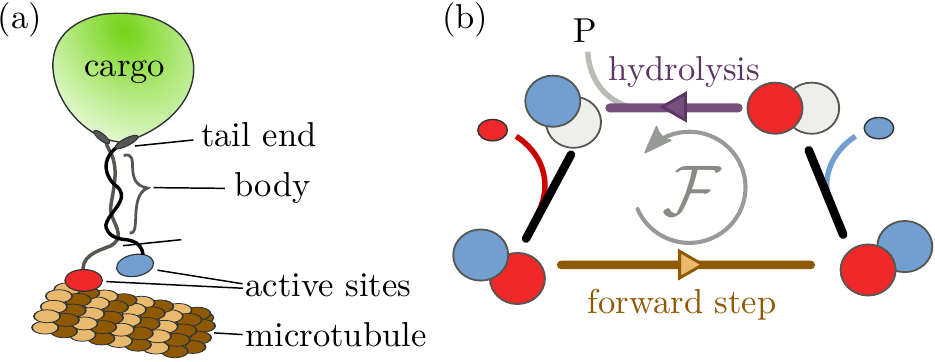}  
 \caption{(a) Kinesin is a motor protein consisting of two identical entangled subunits.
   Cargo is bound at the tail end.
   The active sites on kinesin's head end bind to the microtubule and act as kinesin's ``feet'', enabling the molecule to perform directed steps.
   Kinesin's stepping mechanism is the result of subsequent changes in how strong the active sites bind to the microtubule.
   The trailing (left) and leading (right) sites are represented by colored ellipses.
   ATP-laden (red) and empty (gray) sites bind strongly, whereas an ADP-laden (blue) site binds only weakly.
   The succession of chemical compositions shown in panel (b) is called the forward cycle $\mathcal F$.
   Starting from the upper left state, the forward cycle involves (in counter-clockwise direction):
   (i) Binding of ATP to the (empty) leading site
   (ii) a mechanical step (brown edge), \ie~the exchange of the leading and trailing site,
   (iii) release of ADP from the (new) leading site,
   (iv) hydrolysis (purple edge) of ATP into ADP at the trailing site.
 }
 \label{fig:kinesin-schematic}
\end{figure}
\relax

\subsection{Kinesin as a molecular motor}

Kinesin is a molecular motor which facilitates transport in eukaryotic cells.
It moves along intracellular filaments called microtubules and plays a major role in many biological processes, including mitosis, meiosis and transport of cellular cargo.
The most well studied variety of kinesin -- both experimentally~(see \eg~\cite{Visscher.etal1999,Carter.Cross2005,Ritort2006,Clancy.etal2011} and References therein) and theoretically~\cite{Lau.etal2007,Lacoste.etal2008,Liepelt+Lipowsky2007,Liepelt+Lipowsky2009,Hyeon.etal2009,Zheng.etal2009} -- is a protein dimer consisting of two identical subunits.
Figure~\ref{fig:kinesin-schematic}a shows a sketch of kinesin binding its intracellular cargo at its tail end.
Kinesin's head end consists of two active sites which bind and unbind to the microtubule in alternating succession, thereby allowing the motor to perform mechanical steps of length $L=\SI{8}{\nano\meter}$~\cite{Svoboda.etal1993,Yildiz_etal2004}.
Due to the polarity of the microtubule, this motion has a preferred ``forward'' direction. 

The energy necessary for this active directed transport is provided by the hydrolysis of adenosine triphosphate (ATP) into adenosine diphosphate (ADP) and inorganic phosphate (P).
Unlike macroscopic motors, small molecular machines operate at low Reynolds numbers and inertia plays no role:
chemical energy is not converted into mechanical energy by a transfer of momentum.
Instead, kinesin's mechanical displacement is the result of a complex interplay of the strength of the microtubule binding at the active sites, which depends on their chemical composition.
ATP-laden and empty sites bind strongly, while ADP-laden sites bind weaker~\cite{Carter.Cross2005,Clancy.etal2011}.
Under physiological conditions the mechano-chemical interaction can be described by the ``forward cycle'' depicted in Fig.~\ref{fig:kinesin-schematic}b \cite{Liepelt+Lipowsky2007}.
Models that only treat the forward cycle feature \emph{tight coupling} between the hydrolysis reaction and the stepping:
each hydrolysis of an ATP molecule gives rise to exactly one motor step \cite{Schnitzer.Block1997}.

\subsection{Experiments and models}

An important biophysical question regards the force that kinesin generates for different concentrations of the chemicals $\text{ATP}$, $\text{ADP}$ and $\text{P}$ involved in the hydrolysis reaction.
Typically, experiments measure this force by linking kinesin to a di-electric colloidal bead which resides in an optical trap~\cite{Svoboda.etal1993,Schnitzer.Block1997,Visscher.etal1999,Carter.Cross2005}.
Involved experimental set-ups allow the precise control of the pulling force $F$ that the optical trap exerts on the motor against its typical direction of motion. 
The independent driving parameters are the non-dimensionalized force $f := (L F)/(\kb T)$ and the non-dimensionalized chemical potential difference
\(\Delta \mu = \log\left({K_{\text{eq}}[\text{ATP}]}/{[\text{ADP}][\text{P}]}\right)\),
where $\kb$ denotes Boltzmann's constant, $T$ the temperature, $K\tsub{eq}$ the equilibrium constant of the hydrolysis reaction and $[\text{X}]$ the concentration of chemical species X.
In the remainder of this work, all physical quantities are expressed in units based on the length scale $L$, time scale \SI{1}{\second}, and energy scale $k\tsub{B} T$.

Many experiments probe the stalling force $f\tsub{stall}(\Delta\mu)$, which is defined as the value of the force needed to bring the motor to a halt for a given chemical potential difference $\Delta\mu$.
Under physiological chemical conditions, kinesin hydrolyzes ATP even at stalling forces~\cite{Visscher.etal1999,Carter.Cross2005}.
The exact details of the kinesin stepping mechanism under high mechanical loads remain unknown and several models exist, \cf~Refs.~\cite{Lau.etal2007,Liepelt+Lipowsky2007,Clancy.etal2011} and the references discussed in these publications. 
While these models differ in their details, they all feature more than only the tightly coupled forward cycle. 

A prominent example of a thermodynamically consistent model was introduced in Ref.~\cite{Liepelt+Lipowsky2007}.
There, the key idea is to extend the forward cycle shown in \Fig{kinesin-schematic}b by the chemical compositions obtained from exchanging the trailing with the leading active sites, \cf~\Fig{kinesin6state-cycles}a.
In addition to the forward cycle $\mathcal F$, the extended network features six states and has two additional cycles, Figs.~\ref{fig:kinesin-schematic}b--d:
the backward cycle $\mathcal{B}$ represents backward motion under hydrolysis, while the dissipative slip cycle $\mathcal{D}$ represents the futile hydrolysis of two ATP molecules without any stepping \cite{Liepelt+Lipowsky2009}.
In such a multiple-cycle model, hydrolysis and mechanical displacement are not tightly coupled anymore.
In Sec.~\ref{sec:qtc} we will address the question of \emph{quasi-tight coupling}, \ie~situations where the ratio of the average number of chemical and mechanical events predicted by the model is close to unity.

\subsection{Network theory for the kinesin model}

\begin{figure}[t]
  \centering
  \includegraphics[width=.48\textwidth]{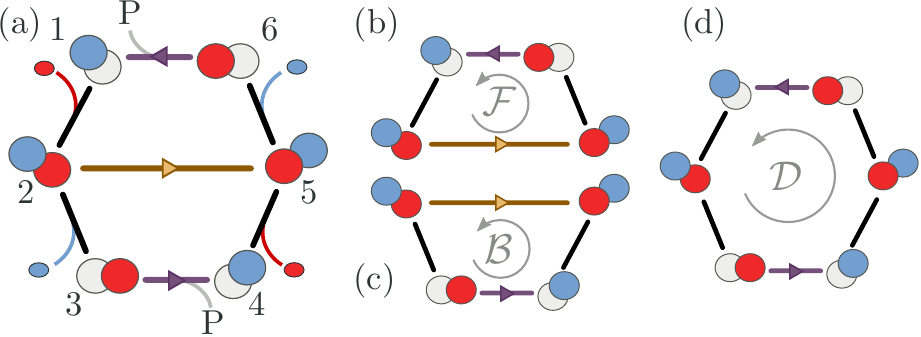}
  \caption{(a) The six-state kinesin model from Ref.~\cite{Liepelt+Lipowsky2007} extends the forward cycle from \Fig{kinesin-schematic}b by two states.
  For this model, We use the same spanning tree and chords as in the example shown in \Fig{trees-cycles}c.
  Then, the fundamental cycles $\zeta_1 \equiv \mathcal{F}$ and $\zeta_2\equiv\mathcal{D}$ are the forward and dissipative cycle (b) and (d), respectively.
  The backward cycle (c) is the linear combination $\mathcal B = \zeta_2-\zeta_1$, \cf~Ref.~\cite{WachtelVollmerAltaner2015}.}
  \label{fig:kinesin6state-cycles}
\end{figure}


In order to study quasi-tight coupling, energy conversion and the predicted response to changes in the driving parameters, we apply the algorithm presented in Sec.~\ref{sec:algorithm} to the six-state model for kinesin, \Fig{kinesin6state-cycles}.
For the first step of the algorithm, we choose the spanning tree and its chords in the same way as in \Fig{trees-cycles}c,d.
Consequently, the fundamental cycles $\zeta_1\equiv\mathcal F$ and $\zeta_2\equiv\mathcal D$ correspond to the forward and dissipative cycles, respectively.

The second step of the algorithm requires the determination of the fundamental current vector $\vec{c}$ and the fundamental co-variance matrix $\vec{C}$.
With the enumeration of the vertices as in \Fig{kinesin6state-cycles}a the matrix \(\tmat_\chords(q_1,q_2)\) reads:\\
\begin{align*}
  \begin{pmatrix}
    w^1_1 & w^1_2 & 0 & 0 & 0 & w^1_6 \\
    w^2_1 & w^2_2& w^2_3 & 0 & w^2_5\,e^{q_1} & 0  \\
    0 & w^3_2 &w^3_3 & w^3_4 &0 & 0 \\
    0 & 0 & w^4_3 &  w^4_4 &  w^4_5 e^{q_2} & 0 \\
    0 & w^5_2 e^{-q_1} & 0 & w^5_4 e^{-q_2} & w^5_5 &  w^5_6\\
    w^6_1 & 0 & 0 & 0 & w^6_5   &w^6_6
  \end{pmatrix}.
\end{align*}
It is straightforward to write down its characteristic polynomial \mbox{\(\chi_\chords(\lambda;q_1,q_2)=: \sum_{k=0}^6 a_k(q_1,q_2)\lambda^k\)}, and to extract the coefficients \(a_0(q) \equiv \det{\tmat_\chords(q_1,q_2)} \), \(a_1\) and \(a_2\).
Differentiating with respect to \(q_1\) and \(q_2\), and evaluating at \(q_1=q_2=0\) yields the expressions $\del_\ell a_k$ appearing in Eqs.~\eqref{eq:cfundamental}.

The third step requires the circulations of the jump observables of interest.
For the present discussion, we consider the displacement $d= \obs_{(2,5)}$ and the hydrolysis count $h=\obs_{(6,1)} + \obs_{(3,4)}$, which indicate a transition along the brown and purple edges in \Fig{kinesin6state-cycles}, respectively. 
Their matrix representations read
\begin{subequations}
\begin{align}
  d^i_j &= \delta_{i,2}\delta_{j,5} - \delta_{i,5}\delta_{j,2},  \\
  h^i_j &= \delta_{i,6}\delta_{j,1} - \delta_{i,1}\delta_{j,6} +  \delta_{i,3}\delta_{j,4} - \delta_{i,4}\delta_{j,3},
\end{align}
\end{subequations}
where $\delta_{m,n}$ denotes the Kronecker delta, which yields one if $m=n$ and zero otherwise.
The circulations of $d$ and $h$ simply count the number of the brown and purple edges in the fundamental cycles $\zeta_1=\mathcal F$ and $\zeta_2 = \mathcal D$, \cf{} \Fig{kinesin6state-cycles}b,d.
The chord representations of $d$ and $h$ thus are
\begin{subequations}
\begin{align}
  d_\chords = (\circu{d}{1},\circu{d}{2}) &= (1,0) \, ,\\
  h_\chords = (\circu{h}{1},\circu{h}{2}) &= (1,2) \, .
\end{align}
\label{eq:chord-representation}
\end{subequations}
Note that the choice of the chords is optimal for the calculation of the present variables because one of the entries of $d_\chords$ vanishes ($\circu{d}{2}=0$), while this cannot be achieved for $h_\chords$. 
After all, all cycles contain at least one hydrolysis event.

The corresponding macroscopic currents, \ie~ the velocity $\ssav{d}$ and the hydrolysis rate $\ssav{h}$ are obtained from \Eq{multilinear-magic} as:
\begin{subequations}
\begin{align}
  \ssav{d}  &= c_1 \, , \\
  \ssav{h}  &= c_1 + 2 c_2 \,.
\end{align}  
  \label{eq:average-currents}
\end{subequations}
Their scaled (co-)variances amount to
\begin{align*}
  \ssav{d,d} &= C_{11},\\
  \ssav{h,h} &= C_{11} + 4C_{12} + 4C_{22},\\
  \ssav{h,d} &= C_{11} + 2C_{12}.
\end{align*}

In addition to displacement $d$ and hydrolysis count $h$, we are interested in the jump observable $\sigma^i_j = \ln(w^i_j/{w^j_i})$ corresponding to the dissipation.
As discussed in Sec.~\ref{sec:jump-observables} its circulations are the cycle affinities.
The Hill--Schnakenberg conditions are necessary for the consistency of a Markov jump process with the thermodynamic notion of local equilibrium \cite{Seifert2008,Esposito.VandenBroeck2010}.
They state that the affinity of a cycle must express the (non-dimensionalized) differences in the potentials of the reservoirs, \cf{} Refs.~\cite{Hill1966,Hill1977,Schnakenberg1976}.
Upon completing the forward cycle $\mathcal F \equiv \zeta_1$, an amount $\Delta\mu$ of chemical energy is used by the system to perform a (dimensionless) amount $-f$ of work against the pulling force.
Similarly, a completion of the dissipative cycle $\mathcal D \equiv \zeta_2$ uses $2\Delta\mu$ of chemical energy.
Consequently, the chord representation of the dissipation reads
\begin{align*}
  \sigma_\chords = (\circu{\sigma}{1},\circu{\sigma}{2}) &= (-f + \Delta\mu,2\Delta\mu).
\end{align*}
The Schnakenberg decomposition thus lets us express the average steady-state dissipation by physical parameters and currents through fundamental chords 
\begin{align}
  \ssav{\sigma}  &= (-f + \Delta\mu) c_1 + (2\Delta\mu) c_2 \label{eq:kinesin-schnakenberg}.
 \end{align}

\subsection{The cycle perspective}
\label{sec:cycles}
\begin{figure}[t]
  \centering
  \includegraphics[width=.48\textwidth]{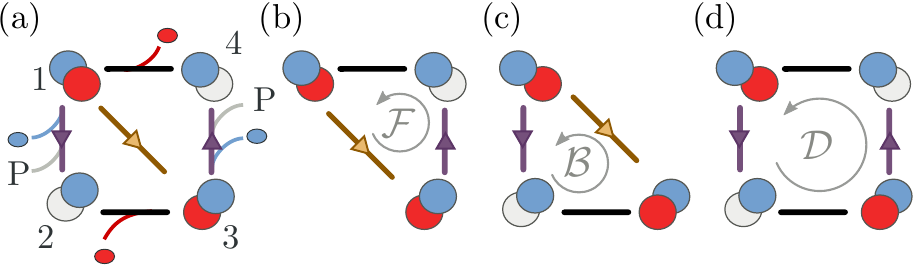}
  \caption{A model with four states, which describes the same physics as the six-state model shown in \Fig{kinesin6state-cycles}.
    In this simpler model we combined the transitions for the ATP hydrolysis on one active site with the ADP release on the other one into a single transition.
    Details of the model construction are given in App.~\ref{sec:construction}.
}
  \label{fig:kinesin4state-cycles}
\end{figure}

%
%

In the previous section, we expressed observable quantities only by means of their circulations around fundamental cycles and the fundamental first and second chord cumulants.
Nowhere in these expressions do the number of states or the choice of a spanning tree appear explicitly.
Hence, the same expressions are reproduced by any model with the same cycle topology -- as long as the physics along the cycles, \ie{} the circulations of antisymmetric jump observables, are the same.
As exemplified in Sec~\ref{sec:jump-observables} in \Fig{trees-cycles}a,b, one can formulate a model on four states with the same cycle topology as the six-state model described in \Fig{kinesin6state-cycles}.
Allowing only for single edges between states, such a four-state model is the minimal model featuring two independent cycles.
An interesting question is how this model (and other reduced models) compare to more complicated ones.

In Ref.~\cite{Altaner.Vollmer2012} we used the idea of preserving the cycle affinities and circulations of jump observables along cycles (together with locality constraints) to develop a coarse-graining algorithm for stochastic models.
Its application to the six-state kinesin model produced various topologically equivalent models, which all preserved the fluctuation statistics of the observables of interest almost perfectly.
A disadvantage of this coarse-graining algorithm is that the individual transitions in the network of states lose their original interpretation.

In contrast, Fig.~\ref{fig:kinesin4state-cycles} shows a four-state model with a clear interpretation of the transitions.
This has the advantage that the parameterization of the transition rates is found by the same physical arguments as the ones used in Ref.~\cite{Liepelt+Lipowsky2007} for the six-state model.
Details of the construction of this model are given in Appendix~\ref{sec:construction}.
We will see in the next section that all the predictions of the six-state model are also found in topologically equivalent four-state models. 
This observation underlines the virtue of viewing models in ST from the perspective of cycles  -- an idea that was pioneered by Hill~\cite{Hill1966,Hill1977} and Schnakenberg~\cite{Schnakenberg1976} and has regained considerable attention recently~\cite{Qian2005,Liepelt+Lipowsky2007,Liepelt+Lipowsky2009,Altaner.etal2012,Altaner.Vollmer2012}.

\section{Results}
\label{sec:results}

\begin{figure*}[t!]
\centering
\includegraphics[scale=1]{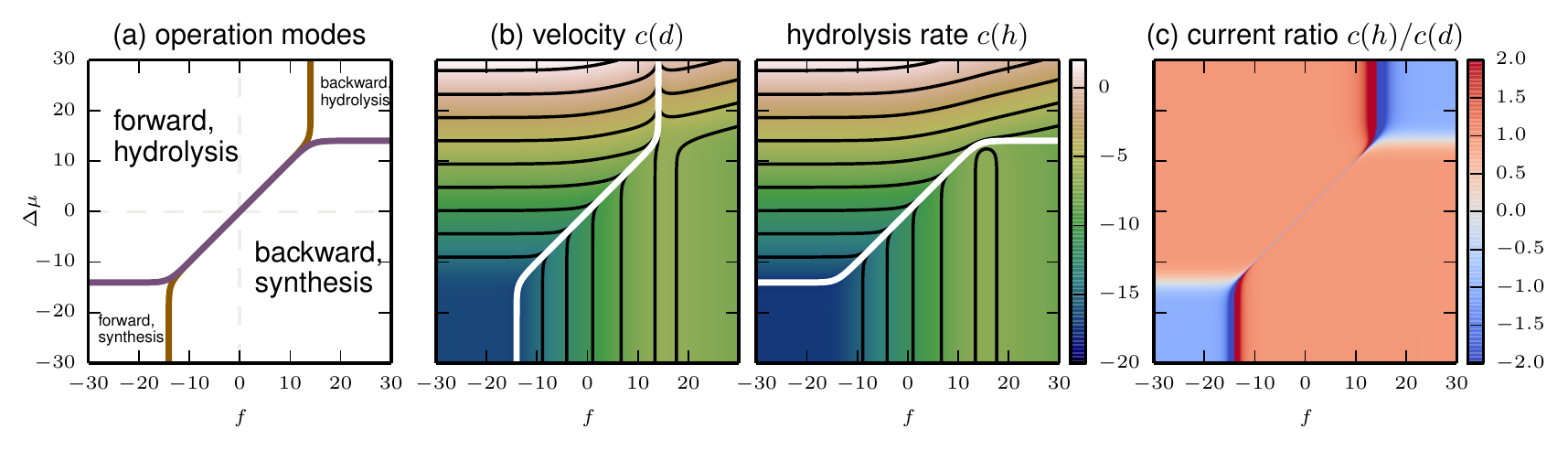}%
 \caption{(a) Operation modes as identified in Ref.~\cite{Liepelt+Lipowsky2009}.
 (b) Decadic logarithm of the absolute value of the average velocity $\ssav{d}$ (left) and hydrolysis rate $\ssav{h}$ (right).
   The white lines indicate where the currents vanish, \ie{} the solid lines displayed in panel (a).
   The contour lines show that the macroscopic currents are proportional away from these lines.
   (c) Plotting the ratio \(\ssav{h}/\ssav{d} \) makes this proportionality visible directly.
   The proportionality constant has an absolute value very close to unity, indicating quasi-tight coupling for most parameter values (see discussion in the main text).
   Note thate the values of the ratio are cropped at absolute values of \num{2}, most prominent in the dark regions surrounding the singularities of the ratio.
 }
 \label{fig:currents-qtc}
\end{figure*}
\relax

One of the main messages of this work is that the algorithm presented in Sec.~\ref{sec:algorithm} allows us to probe parametric models used in Stochastic Thermodynamics in a systematic way.
In order to demonstrate the efficiency of our approach, we report on various non-trivial predictions of the six-state kinesin model.
At the end of this section we will compare these results with other models.

Throughout this section, we choose the parameter range similar to Ref.~\cite{Liepelt+Lipowsky2009} and vary  $-30 \leq f,\chempot \leq 30$.
Then, in physical units the pulling force $F$ varies between about $-15$ and \(\SI{+15}{\pico\newton}\).
Following Ref.~\cite{Liepelt+Lipowsky2007}, the chemical potential difference is adjusted by changing the ATP concentration while fixing the other chemical concentrations at physiological values (see also Appendix \ref{sec:construction}).
The physiologically relevant region for the chemical driving parameter is limited to about $20<\chempot<30$.
Negative values of the chemical potential correspond to extremely low ATP concentrations.
In particular, the ``homeopathic limit'' is reached at about $\chempot < -14$:
At that point there is less than one ATP molecule in an experiment containing one liter of solution.

The reason we still present our results for the entire parameter range $-30 \leq f,\chempot \leq 30$ is two-fold:
firstly, it enables a direct comparison with previous work \cite{Liepelt+Lipowsky2007,Liepelt+Lipowsky2009}.
Secondly, it demonstrates the effectiveness of our algorithm in predicting results that vary over many orders of magnitude.
Still, we emphasize that  non-trivial results are encountered exactly in the experimentally accessible region, where we consider the model as valid.

\subsection{Velocity, hydrolysis rate, and their quasi tight coupling}
\label{sec:qtc}
\Fig{currents-qtc}a reproduces a central result of Ref.~\cite{Liepelt+Lipowsky2009} concerning the operations modes of kinesin.
These modes are defined by the signs of the average currents, \ie{} of the velocity \(\ssav{d}\) and the hydrolysis rate \(\ssav{h}\).
However, the resulting phase diagram contains no information regarding their magnitude.
Based on the expressions \eqref{eq:average-currents} we provide a detailed account on their numerical values in \Fig{currents-qtc}b.
Note that these currents vary over about \num{20} orders of magnitude.
This underlines the importance of having analytical expressions to generate the plots.
A brute-force numerical approach will either be prohibitively expensive in terms of computer resources, or it will suffer from severe inaccuracies when dealing with this vast range of numerical values. 

The analytical expressions for the currents also reveal an interesting relation between the average currents.
In \Fig{currents-qtc}d we plot the ratio \(\ssav{h}/\ssav{d}\) of the hydrolysis rate and the velocity.
Again, access to the analytical expressions for the currents is crucial to determine the ratio.
After all, both its numerator and denominator are of the order \num{e-18} in the lower left corner of the parameter space.

The most prominent feature of \Fig{currents-qtc}c is that away from the zero-current lines, the ratio of average hydrolysis rate and velocity takes values very close to \num{\pm 1}. 
Consequently, on average the completion of a cycle yields one mechanical step and one chemical event.
We say that chemical and mechanical currents are \emph{quasi-tightly coupled}.
Experimentally, it was found that kinesin hydrolyzes one ATP molecule for each mechanical step~\cite{Schnitzer.Block1997}.
According to the model considered here, quasi-tight coupling is a generic feature that holds more generally:
even in the region where kinesin moves \emph{backward} while consuming ATP, the absolute values of the currents are locked to a ratio of one.

Knowing the absolute values of the currents rather than only their signs also allows us to treat kinesin's thermodynamic cycles in more detail.
In Ref.~\cite{Liepelt+Lipowsky2009} this discussion was based on the signs of Hill's (excess) cycle fluxes~\cite{Hill1966}.
With the current ratio we interpret the regions shown in the phase diagram \Fig{currents-qtc}a in terms of dominant cycles -- at least away from their boundaries:
in the upper left and lower right regions the forward cycle $\mathcal{F}$ dominates such that average hydrolysis and velocity are directly proportional, \(\ssav{h}\sim \ssav{d}\).
The difference between those regions is the angular direction: 
counter-clockwise completion leads to a forward movement accompanied by ATP hydrolysis, whereas clockwise completion yields backward stepping and ATP synthesis. 
In contrast, in the upper right and lower left regions hydrolysis and velocity are anti-proportional, \(\ssav{h}\sim ~-\ssav{d}\):
a counter-clockwise (backward, hydrolysis) or a clockwise (forward, synthesis) completion of the backward cycle $\mathcal{B}$ dominates the average dynamics, respectively.
This result, which is based on the values of physiological currents, thus complements and extends the discussion presented in Ref.~\cite{Liepelt+Lipowsky2009}.

\subsection{Efficiency of energy conversion}
\label{sec:efficiency}

Under physiological conditions, kinesin uses the chemical energy released by the ATP hydrolysis to perform mechanical work.
Energy efficiency is one of the most important questions for molecular machines involved in cellular energy conversion~\cite{Esposito.etal2009,Verley.etal2014,Polettini.etal2015}, just as it is for macroscopic machines.
A framework for a quantitative analysis is based on the notion of conjugate currents and forces from irreversible thermodynamics~\cite{Groot1984}.
Generally, a complete set of conjugate currents $\ssav{\obs_i}$ and forces ${E_i}$ yields the average dissipation as the bilinear form
\begin{align*}
  \ssav{\sigma} = \sum_i \ssav{\obs_i}E_i =: \sum_i \ssav{\sigma_i},
\end{align*}
where $\sigma_i=\obs_i E_i$ denotes the distinct contributions to the entropy production.

Using equations \eqref{eq:average-currents} and \eqref{eq:kinesin-schnakenberg} we find that
\begin{align}
  \ssav{\sigma} = (-f)\ssav{d} + (\Delta\mu)\ssav{h} \nonumber \\=: \ssav{\sigma\tsub{mech}} + \ssav{\sigma\tsub{chem}}.
  \label{eq:kinesin-conjugate}
\end{align}
We see that in the kinesin model, velocity $\ssav{d}$ and hydrolysis rate $\ssav{h}$ are conjugate to the negative pulling force $-f$ and the chemical potential $\Delta\mu$, respectively.
For a system with two independent contributions to the entropy production, \(\sigma = \sigma_1 +\sigma_2\), one may define the efficiency of energy conversion in general terms~\cite{Verley.etal2014}.
To that end note that $\ssav{\sigma}$ is always positive.
This, however, does not imply that both contributions \(\ssav{\sigma_i}\) are positive.
Indeed, systems act as energy converters only if one of the contributions, say $\sigma_1$, is negative.
Then, a (positive) average power output \(\dot{W}\tsub{out} :=-\ssav{\sigma_1} \) is sustained by a (positive) average power input \(\dot{W}\tsub{in}=\ssav{\sigma_2}\).
Note that \(\ssav{\sigma_2}\) is positive and larger in magnitude than $\ssav{\sigma_1}$, because $\ssav{\sigma} = \ssav{\sigma_1} + \ssav{\sigma_2} \geq 0$ must always hold. 
Hence, the efficiency of energy conversion is defined as
\begin{align}
  0\leq\hat\eta \coloneqq \frac{\dot{W}\tsub{out}}{\dot{W}\tsub{in}} = \frac{|\sigma_1|}{\sigma_2} < 1 \, .
  \label{eq:efficiency}
\end{align}
It is always positive and smaller than unity.

\begin{figure}
\centering
 \hspace*{\fill}  
 \includegraphics[width=.48\textwidth]{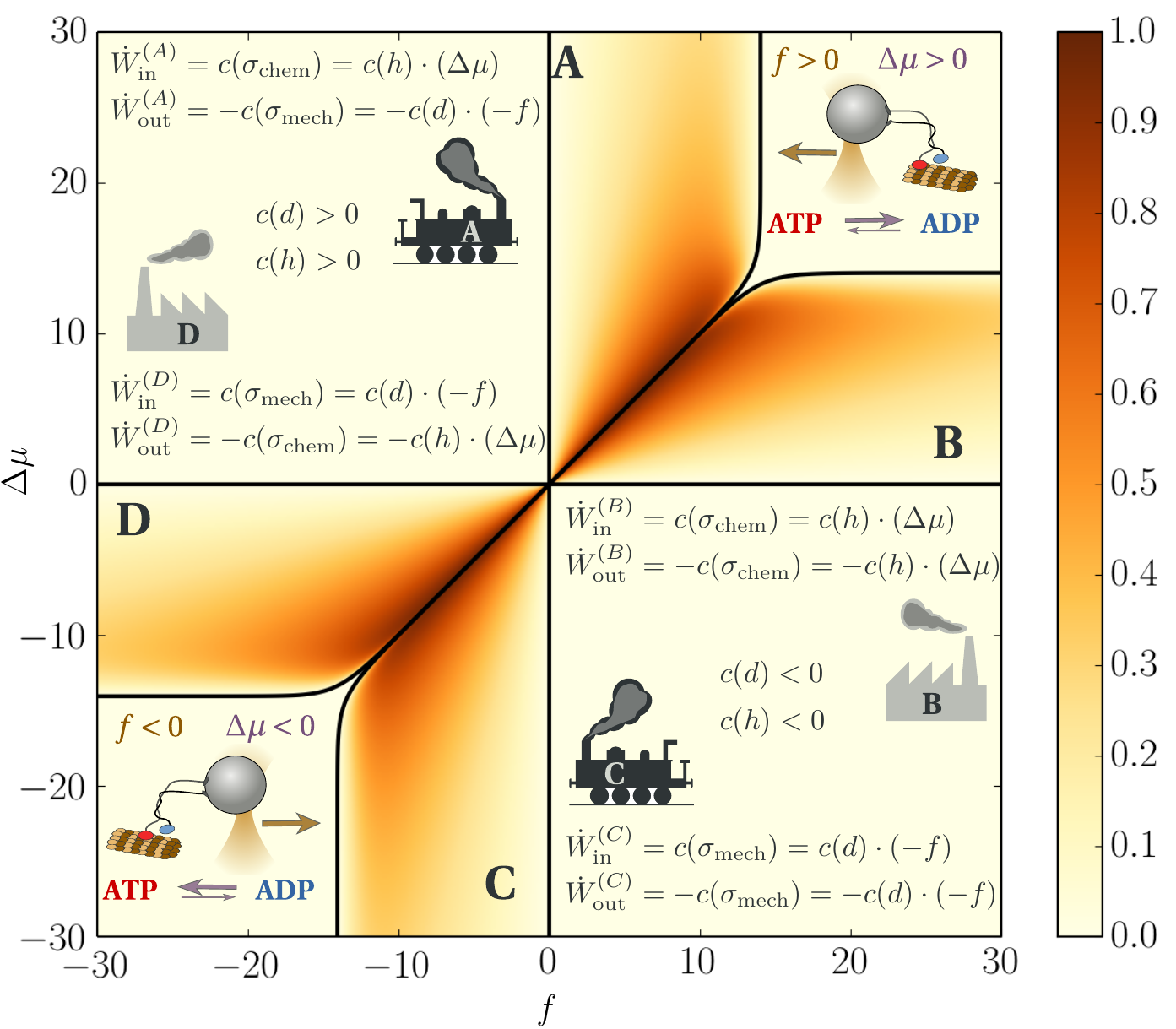}%
  \caption{Efficiency of energy conversion in the kinesin model.
    The four regions A--D correspond to four different ways of energy conversion.
    In the regions outside the solid curves, no conversion between mechanical and chemical energy takes place.
    In regions A and C kinesin acts as a motor converting chemical into mechanical energy against the external force.
    In regions B and D kinesin resembles a chemical factory that uses mechanical energy to produce ATP and ADP, respectively, against the chemical potential provided by the solution.
    The sketches in the upper right and lower left illustrate the combination of thermodynamic forces acting on the motor in the respective quadrant.
    In the upper right, kinesin is pulled backward (\ie{} against its standard direction of motion) in an ATP rich environment.
    In the lower left, kinesin is pushed forward in an ADP rich environment.
    Energy conversion only occurs in the regions where both mechanical and chemical currents have the same sign and the forward cycle dominates, \cf{} \Fig{currents-qtc}.
 }
 \label{fig:efficiency}
\end{figure}

In the framework of Stochastic Thermodynamics, this efficiency has been studied under various aspects (\cf{} \eg{}~Refs.~\cite{Seifert.Speck2010,Esposito.etal2009,Verley.etal2014,Polettini.etal2015}).
In \Fig{efficiency}, we give the efficiency of energy conversion $\hat\eta$ for the kinesin model.
The regions A--D correspond to different types of energy conversion where the system either acts as a motor (A,C) or a chemical factory (B,D).
Outside of these regions both contributions to the entropy production are positive and no energy conversion takes place.

We note the following prediction:
for any fixed value of $\chempot$ in the physiological range, \ie{} for \(20< \chempot <30\), the value of the force at maximum efficiency is around \(f\approx 10.5\). 
This suggest that kinesin might be optimized to encounter (elastic) forces of around \(\SI{5}{\pico\newton}\), independent of the ATP concentration.
It will be interesting to explore the implications of this result for the collaborative behavior of multiple kinesin molecules involved in the viscous transport of organelles.

\revi{
Finding the parameters of a system that extremize thermodynamic quantities is a generic problem.
Recently, many authors have discussed the notion of efficiency at maximum power (see \eg{} Refs.~\cite{Esposito.etal2009,Seifert2011a} and therein).
Having fully parameter-dependent symbolic expressions for the various contributions to the entropy production establishes a general (analytic) approach to this optimization problem.
}

\subsection{Diffusion constant and randomness parameter}
\label{sec:diffusion}
\begin{figure}
 \hspace*{\fill} 
   \includegraphics[width=.5\textwidth]{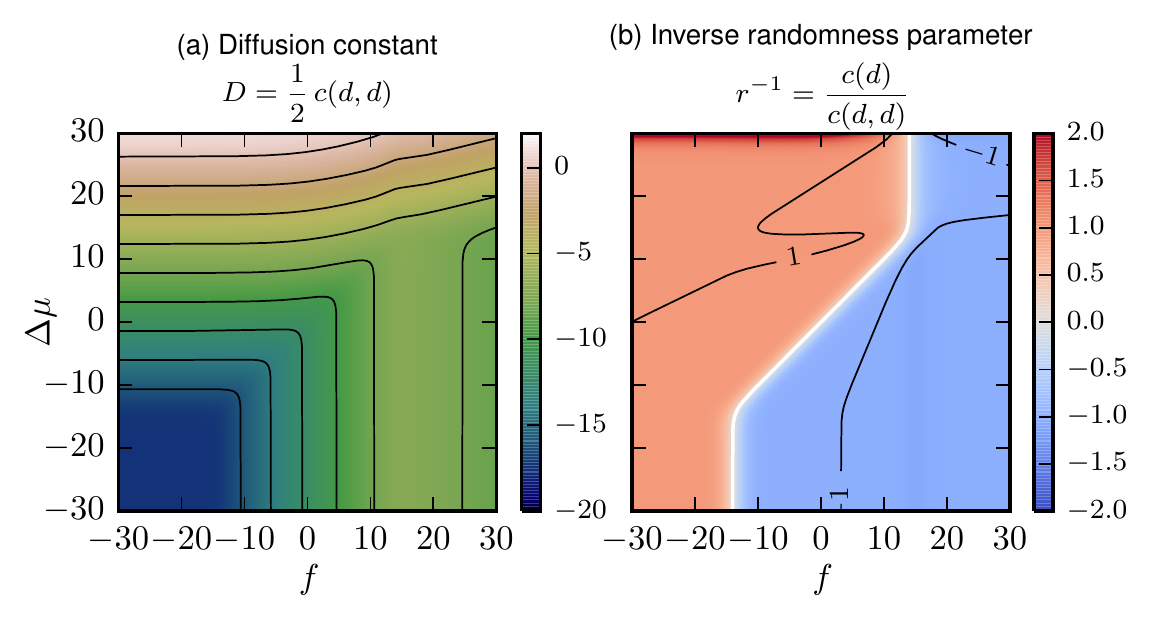}%
 \hspace*{\fill}%
 \caption{
   (a) The diffusion constant \(D = \frac{1}{2}c(d,d)\) on a decadic logarithmic scale.
   (b) The inverse randomness parameter \(r^{-1}= c(d)/c(d,d) \equiv \ssav{d}/(2D)\) compares the velocity and the diffusion constant. 
   Away from the stalling line, it obtains an absolute value close to unity.
   Solid lines show where $\abs{r} =1$ holds exactly. 
 }
 \label{fig:diffusion}
\end{figure}
So far, we have only investigated average currents, which are also available if the steady-state distribution $\vec{\pi}$ is known, \cf~\Eq{macro-current}.
Higher order statistics of fluctuating currents cannot be expressed by means of the stationary distribution only, although a perturbation expansion exist~\citep{2009JoSP-Baiesi.etal}.
The method presented here provides direct access to the (co-)variance of fluctuating currents via \Eq{multilinear-magic-2}, without the knowledge of the stationary distribution.

For motor proteins we are mostly interested in the second scaled cumulant of the time-averaged displacement.
It quantifies the (linear) scaling of the (fluctuating part of) the mean-square displacement, and thus defines (up to a factor of two) the non-equilibrium diffusion constant $D=D(f,\chempot)$:
\begin{align*}
 c(d,d)
 &\equiv\lim_{T\to\infty} \frac{1}{T}\left(\trav{d(T)^2} - \trav{d(T)}^2\right)  =: 2D.
\end{align*}

In \Fig{diffusion}a we show the diffusion constant in the six-state kinesin model.
Like the average velocity, its values span a range of about 20 orders of magnitude.
Under physiological conditions (\(f=0, \chempot \approx 25\)) the diffusion constant is about ten orders of magnitude larger than at equilibrium.

A direct measurement of the parameter dependence of $D$ is difficult.
An observable that is more easily accessible in experiments is the so-called \emph{randomness parameter} (sometimes called Fano factor) \cite{Svoboda.etal1994,Visscher.etal1999,Chemla_etal2008,Clancy.etal2011} 
\begin{align}
r=\lim_{T\to\infty}\frac{ \trav{d(T)^2} - \trav{d(T)}^2   }{\trav{d(T)}} = \frac{c(d,d)}{c(d)}\,.
  \label{eq:randomness-parameter}
\end{align}
It is a dimensionless measure of the temporal irregularity of the mechanical displacement.
While \(r=0\) indicates a deterministic motion without any fluctuations, a value of \(\abs{r}=1\) amounts to a Poisson motor \cite{Svoboda.etal1994}.
In \Fig{diffusion}b we plot its inverse, \(r^{-1}\), which is a smooth function.
We see that the six-state model predicts Poissonian behavior \(\abs{r} \approx 1\) in a large area away from the stalling lines.
This is in agreement with recent experimental results and theoretical predictions from an alternative model~\cite{Clancy.etal2011}.

Our method to calculate the second scaled cumulant and thus the diffusion constant avoids all of the combinatorial complexity of previous approaches~\citep{Chemla_etal2008,Boon2012}.
Ref.~\citep{Koza1999} treats drift velocity and diffusion in Markovian models formulated for a periodic lattice in arbitrary dimensions.
In the present work the topology of physical space is independent from the structure of the graph, which represents the topology of the model:
if a system like a molecular motor moves in more than one spatial dimension, one defines a distinct jump observable \(d_i\) for each of these dimensions \(i\).
Up to a factor of two, the scaled covariance matrix \(c(d_i,d_j)\) then equals the diffusion tensor.

\subsection{Response theory}
\label{sec:response}

\Eq{kinesin-conjugate} states that the average velocity $\ssav{d}$ and hydrolysis rate $\ssav{h}$ are conjugate to the mechanical and chemical driving forces $-f$ and $\chempot$.
Response theory studies the dependence of averaged currents $\vec{J} = \left(\ssav{\obs_i}\right)_i$ to the conjugate fields $\vec{E}=\left(E_i\right)_i$.
For $B$ independent conjugate current--field pairs, $(\ssav{\obs_i},E_i)_i$, the response matrix $\boldsymbol{\sfmat{R}}(\vec E)$ is a $B\times B$ matrix with entries
\begin{align}
  \sfmat{R}_{i,j}(\vec{E}) := \left.\frac{\partial\ssav{\obs_i}}{\partial{E_j}}\right\vert_{\vec E}. 
  \label{eq:response-matrix}
\end{align}

Fluctuation dissipation relations (FDR) relate the response of average currents to their fluctuation statistics.
In particular, the Einstein relation relates the mobility of a particle (or its inverse, the friction coefficient) to its diffusion constant.
So called Green--Kubo relations~\cite{Helfand1960} express equilibrium transport coefficients by time-correlation integrals, \Eq{correlation}.
Here, these time-correlation integrals are obtained as second-order scaled cumulants $\ssav{\obs_i,\obs_j}$.
The fluctuation relation for the entropy production~\cite{Maes2004,Seifert2012} ensures the validity of the following equilibrium FDR~\cite{Lebowitz+Spohn1999,Andrieux2007}:
\begin{align}
  \sfmat{R}_{i,j}(\vec E = \vec 0) = \frac{1}{2} \left.c(\obs_i,\obs_j)\right\vert_{\vec{E}=\vec 0}.
  \label{eq:FDR}
\end{align}

With analytical expressions for the average currents $\ssav{\obs_i}$ one can calculate their derivatives $\sfmat{R}_{ij}$.
Because the correlation integrals $c(\obs_i,\obs_j)$ are known, our method enables us to probe the non-equilibrium response properties predicted by models of Stochastic Thermodynamics.
As an example, we discuss kinesin's mechanical response using the normalized response coefficient
\begin{align}
  \sfmat{T}\tsub{mech}(f,\chempot) := \frac{2}{c(d,d)} \frac{\partial c(d)}{\partial{(-f)}} = -\frac{1}{D}\frac{\partial\ssav{d}}{\partial{f}}.
  \label{eq:normalized-response-kinesin}
\end{align}
The equilibrium FDR \eqref{eq:FDR} implies that $\sfmat{T}\tsub{mech}(0,0) =1 $.
As the transition matrix depends smoothly on the driving parameters $(-f,\chempot)$ \cite{Liepelt+Lipowsky2007}, we expect that there will be a one-dimensional curve in parameter space where  $\sfmat{T}\tsub{mech}(f,\chempot) =1 $.
\begin{figure}[t]
  \centering
  \includegraphics[width=.48\textwidth]{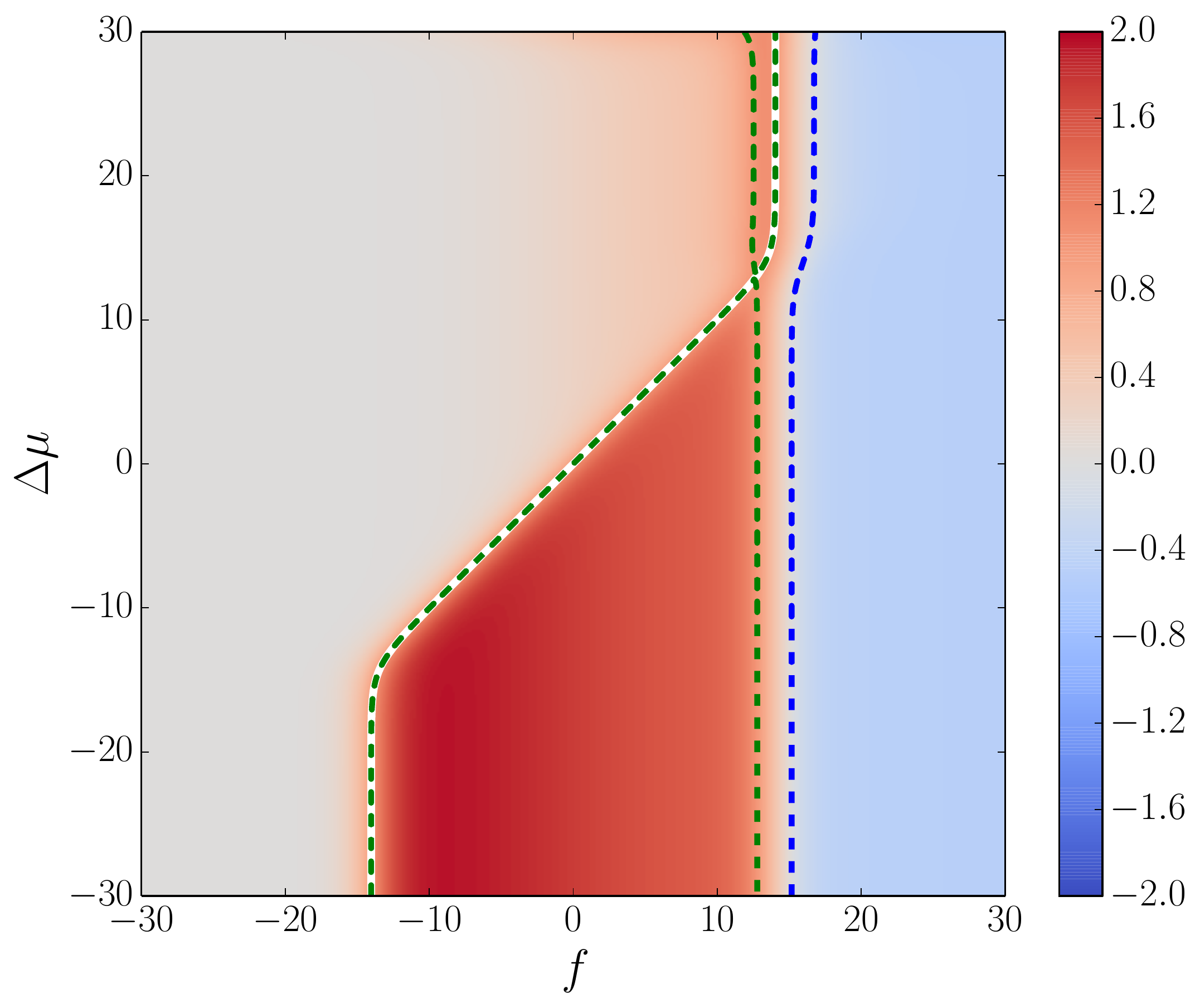}
  \caption{Normalized mechanical response $\sfmat{T}\tsub{mech}$.
  On the dashed green curves, a Green--Kubo FDR \eqref{eq:FDR} holds.
  One of these curves coincides with the stalling line $f=f\tsub{stall}(\chempot)$ where the average velocity vanishes (white line).
  The dashed blue line indicates a vanishing transport coefficient.
  To its right lies a region of negative differential response.
}
  \label{fig:kinesin-response}
\end{figure}

Figure~\ref{fig:kinesin-response} depicts $\sfmat{T}\tsub{mech}$. 
As expected, we see that $\sfmat{T}\tsub{mech}(f,\chempot) =1 $ holds along two lines originating from the origin, such that a nonequilibrium FDR holds for these parameter values.
Remarkably, one of these lines coincides with the stalling line $f = f\tsub{stall}(\chempot)$, \ie{} for parameters where the average velocity vanishes.

Another non-trivial feature of Figure~\ref{fig:kinesin-response} is the region where the normalized mechanical response is negative.
Since the diffusion constant $D$ is positive, $\sfmat{T}\tsub{mech}<0$ implies that the derivative $\partial c(d)/\partial(-f)$ of the mechanical current with respect to its conjugate force is negative.
This phenomenon is known as negative differential mobility \cite{Benichou.etal2014}, or more generally, negative differential response (NDR) \cite{Zia2007,Baerts.etal2013}.
The kinesin model predicts negative differential mobility for large enough pulling forces beyond stalling, \ie{} in situations where the motor walks backwards.
Then, by pulling more one gets less, \ie{} the velocity in pulling direction becomes smaller.
This feature might already be visible in the experimental data found in Refs.~\cite{Carter.Cross2005,Clancy.etal2011}.
Although we do not expect to see NDR for arbitrarily high pulling forces in real experiments, explicitly looking for it in the region for small superstalling forces seems worthwhile.

\subsection{Model comparison}
\label{sec:model-comparison}
%

\begin{figure}[t]
  \centering
  \includegraphics[width=.49\textwidth]{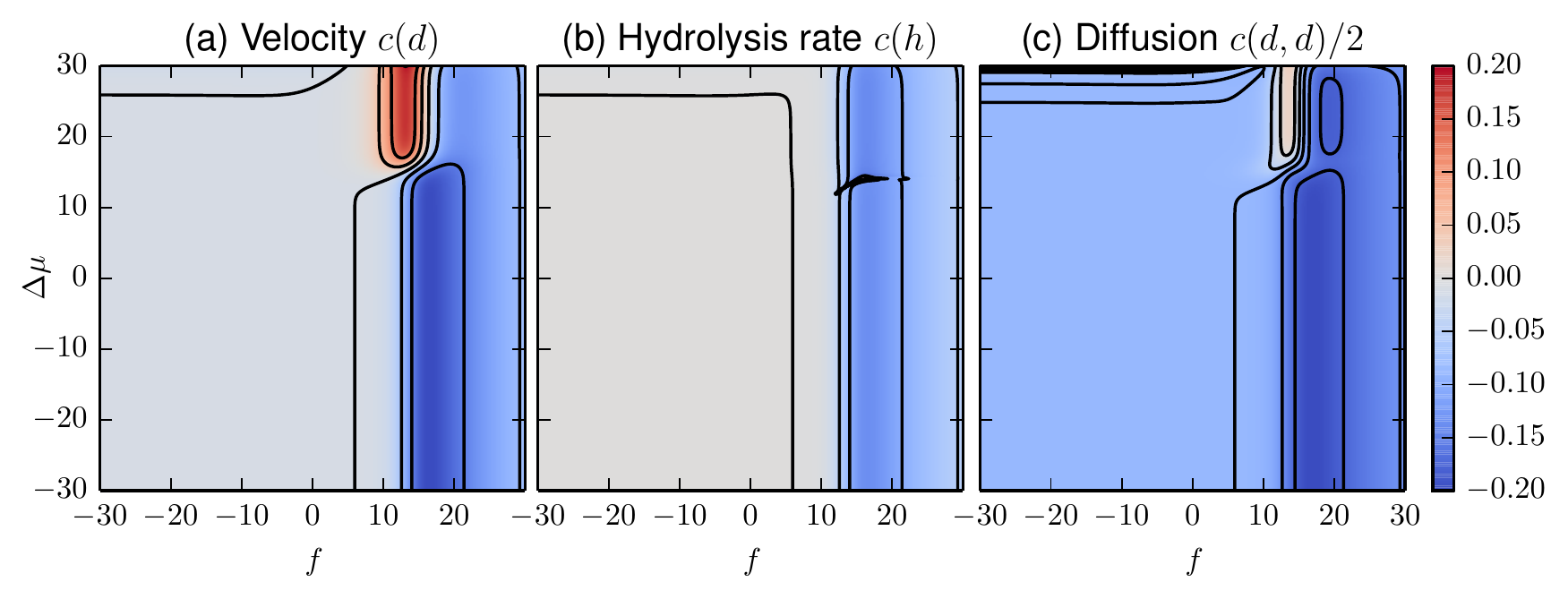}
 \caption{Comparison of our four-state model with the six-state model of Ref.~\cite{Liepelt+Lipowsky2007}.
 We show the relative errors $X_4/X_6 -1$ of the corresponding quantities $X_4$ and $X_6$ calculated in the four- and six-state models, respectively.
 Throughout the parameter range considered, they are almost everywhere well below 15\%.
 Note that for the average hydrolysis rate the relative error diverges close to the line $c(h)=0$ where the hydrolysis rate in the six-state vanishes.
 Remarkably, this is not the case for the average velocity, where the stalling lines in both models agree exactly.
} 
 \label{fig:mismatch}
\end{figure}

Direct access to the non-trivial features of physical currents allows us to compare different models in detail, both qualitatively and quantitatively.
We start by quantitatively comparing  the results for the six-state model (\Fig{kinesin6state-cycles}) with the simpler four-state model (\Fig{kinesin4state-cycles}).
Recall that the latter is constructed following the same physical arguments as the former (\cf{} Appendix \ref{sec:construction} for the details).
The results plotted in Figs.~\ref{fig:currents-qtc}--\ref{fig:kinesin-response} are all derived from $\ssav{d}$, $\ssav{h}$ and $c(d,d)=2D$.
In \Fig{mismatch} we plot the relative deviations of these quantities between the four-state and the six-state model.
Throughout most of the parameter space, they are only a few percent.
This is remarkable, because the observables themselves vary over many orders of magnitude.
Note that at the boundaries between different operation modes (\Fig{currents-qtc}a), the first cumulants vanish.
Hence, we have a divergence in the relative errors unless this happens exactly at the same parameter values in both models.

For the hydrolysis rate $\ssav{h}$ such a divergence is visible in \Fig{mismatch}b around $(f,\chempot)\simeq(16,14)$.
In principle, this divergence is present wherever $\ssav{h}$ vanishes in the six-state model.
In practice, however, the curves of zero average hydrolysis rate $c(h)=0$ agree almost perfectly, such that the region where the divergence has an effect is extremely small.
For most parameters it is hidden due to the finite plotting resolutions, and thus not visible in \Fig{mismatch}b.
In contrast, the prediction of the stalling forces $f\tsub{stall}(\chempot)$ agrees exactly between the two models:
\Fig{mismatch}a does not exhibit any singularities.
In Ref.~\cite{Altaner.Vollmer2012} we introduced a coarse-graining procedure which preserves the cycle topology of a model.
By construction, the first cumulants of \emph{all} currents agree between the original and the coarse-grained models.
Moreover, the relative error in the diffusion constant is comparable in magnitude to what we see in  \Fig{mismatch}c.
These \emph{quantitative} results emphasize the value of the cycle perspective introduced in Sec~\ref{sec:cycles}:
In order to construct thermodynamically consistent models, one should think of the physics of cycles rather than focusing only on individual transitions.

Finally we compare the six-state model to a general model for molecular motors presented in Ref.~\cite{Kafri.etal2004}, which was studied in detail in Ref.~\cite{Lau.etal2007}.
Unlike the six-state model studied fo far, that model features only two states, which correspond to a strongly and a weakly bound configuration.
Multiple transition between these two configurations are possible and represent either an active (\ie{} accompanied by a chemical event) or passive displacement along the microtubule.
The cycles of the two models are different both in their topology and their interpretation.
In particular, the two-state model of Refs.~\cite{Kafri.etal2004,Lau.etal2007} has no reference to the ``hand-over-hand'' stepping mechanism of the forward cycle of Ref.~\cite{Liepelt+Lipowsky2007}, depicted in \Fig{kinesin-schematic}b.
Moreover, the two-state model was fitted to the experiments of Refs.~\cite{Schnitzer.Block1997,Visscher.etal1999}, whereas the six-state model used the experimental data from Ref.~\cite{Carter.Cross2005}

Due to the simple structure of the two-state model, an analytical parameter-dependent expression of the scaled cumulant generating function was found in Ref.~\cite{Lau.etal2007}.
Consequently, analytical expression for the scaled cumulants are known and can be compared to the results obtained for the six-state model of Ref.~\cite{Liepelt+Lipowsky2007}, which we used so far.
Due to the different nature of the models, we do not expect their predictions to agree quantitatively.
In particular this is the case for parameter values that are far away from values that are realized in the actual experiments, \ie{} for small (or even negative) chemical potentials $\chempot$, or negative values of the pulling force.
However, for experimentally accessible parameters, it makes sense to look for \emph{qualitative} agreements in the features of the two different kinesin models.

\begin{figure}[t]
  \centering
  \includegraphics[width=.48\textwidth]{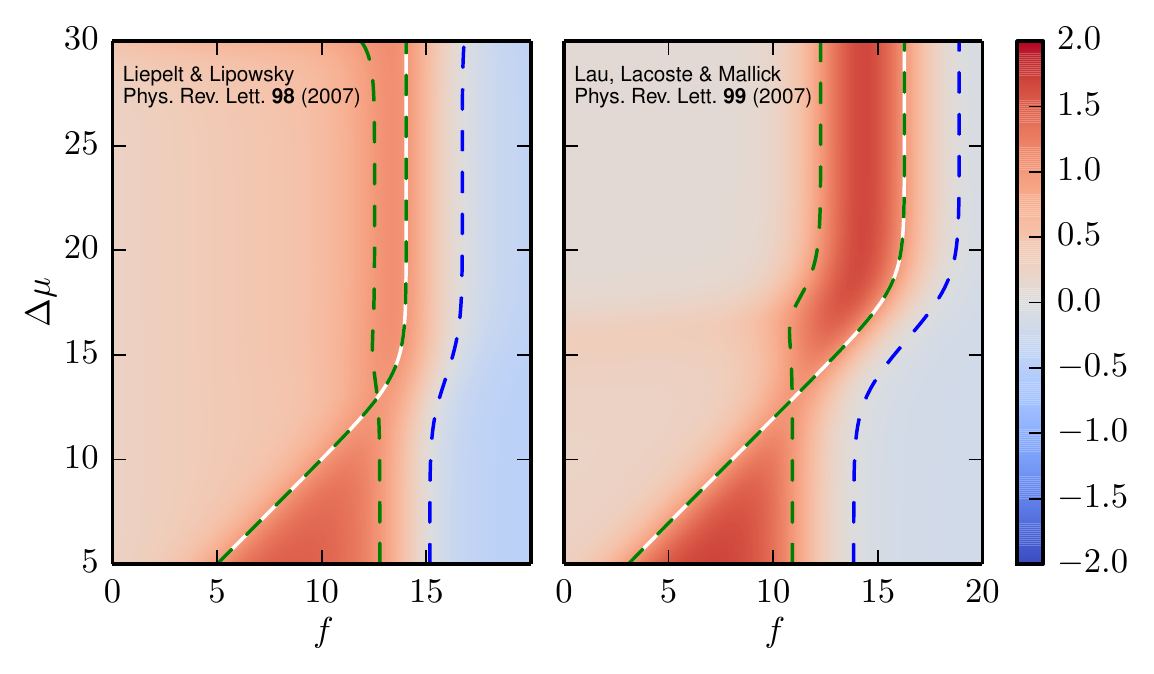}
  \caption{Normalized mechanical response $\sfmat{T}\tsub{mech}$ for the six-state model of Ref.~\cite{Liepelt+Lipowsky2007} studied in the present work (left), and the model from Ref.~\cite{Lau.etal2007} (right).
  For experimentally sensible parameters both models predict the same qualitative behavior:
  The validity of a non-equilibrium Einstein FDR (green curves) at stalling (white curves) as well as negative differential response (regions to the right of blue curves) for superstalling forces.
  Other features (like the overall structure, magnitude of the response) show the same qualitative behavior of the two models.
}
  \label{fig:response-comparison}
\end{figure}
Fig.~\ref{fig:response-comparison} shows the normalized mechanical response Eq.~\eqref{eq:normalized-response-kinesin} in both models for sensible chemical potential differences ($5\leq \chempot \leq 30$) and positive pulling forces.
As expected, the models do not agree quantitatively.
In particular, the stalling lines are at different positions.
However, they show the same qualitative features:
the validity of an Einstein FDR at stalling, and the emergence of negative differential mobility just above stalling.
Together with the experimental hints from Ref.~\cite{Carter.Cross2005}, we consider this agreement as evidence that negative differential mobility is a generic feature of kinesin --- a prediction that should be studied by future experiments.

\section{Conclusion}
\label{sec:conclusion}

In the present work, we gave an explicit procedure for the analytical, \ie{} fully parameter-dependent, calculation of the statistics of fluctuating currents in Stochastic Thermodynamics.
The algorithm applies to any finite Markov model.
We focused on its efficiency in exploring the parameter space of models for the motor protein kinesin, while the mathematical background of the algorithm was the subject of an accompanying paper~\cite{WachtelVollmerAltaner2015}.
In the following we summarize conceptual and physical insights.

From a conceptual point of view, we find the following points particularly noteworthy:
\begin{itemize}
  \item Our algorithm is efficient.
    After obtaining the fundamental chord cumulants, Eqs.~\eqref{eq:multilinear-magic} provide fully parameter-dependent expressions for averages and time-correlations of arbitrary currents.
  \item \revi{Our algorithm is purely symbolic.
    Thus it allows simplification and cancellation of zeroes.} 
    This prevents floating-point inaccuracies even in expressions that vary over many orders of magnitude, \cf{} \Fig{currents-qtc}b or \Fig{diffusion}a.
  \item Having access to symbolic expressions allows further (algebraic) manipulation and thus the study of derived expressions, \cf{} \Fig{currents-qtc}c or \Fig{diffusion}b.
  Taking derivatives with respect to external parameters is needed to explore response properties, \cf{} \Eq{normalized-response-kinesin} \revi{and approach (thermodynamic) optimization problems (\eg{} efficiency at maximum power, \cf{} Sec.~\ref{sec:efficiency})}.
\end{itemize}

From a physical perspective, our method allows the systematic comparison of the predictions made by various kinesin models, \cf{} Sec.~\ref{sec:kinesin}.
In particular, we gave a detailed account on (quasi-)tight coupling, efficiency of energy conversion, diffusion and mechanical response for a well-known kinesin model~\cite{Liepelt+Lipowsky2007} in Secs.~\ref{sec:qtc}--\ref{sec:response}.
Moreover, in Sec.~\ref{sec:model-comparison} we compared these predictions with other models.
Regarding the modeling of molecular motors, we emphasize the following:
\begin{itemize}
  \item Current statistics correspond to experimentally observable quantities, like the average motor velocity or its nonequilibrium diffusion constant.
    Our systematic approach thus extends and unifies previous approaches for calculating these quantities~\cite{Koza1999,Chemla_etal2008,Boon2012}.
  \item Thinking of stochastic models in terms of its physical cycles is useful.
    It allows model reduction, \cf{} Sec.~\ref{sec:model-comparison} and Ref.~\cite{Altaner.Vollmer2012}.
  \item Independent models predict two interesting nonequilibrium response properties of kinesin:
    (i) the validity of a non-equilibrium fluctuation-dissipation relation at mechanical stalling, and (ii) negative differential mobility for superstalling forces.
    Both of these predictions lie in realistic parameter regions and can be tested in future experiments.
\end{itemize}

Modeling the dynamics of molecular motors as random transitions on a biochemical network of states is only one of many appliations of finite Markovian jump processes.
The methods established in the present paper apply to any other dynamically reversible model and are easily extended to systems with multiple transitions between states.
Thus, they provide a powerful framework to fully explore the physical predictions of any model described by Stochastic Thermodynamics.

\subsection*{Acknowledgements}

The authors thank Nigel Goldenfeld, Matteo Polettini and Massimiliano Esposito for insightful discussions.
JV acknowledges a research grant of the  ``Center for Earth System Research and Sustainability'' while the final version of this manuscript was drafted.

\appendix

\section{Transition rates of the four-state model for kinesin}
\label{sec:construction}

\begin{table*}
 \centering
 \begin{tabular}{lr@{\,=\,}lr@{\,=\,}l}
  \toprule Mechanical transition& \(\kappa^1_3\)&
  \num{3e5}&
   \(\kappa^3_1\)&
  \num{0.24}\\
  \midrule Chemical transitions&
   \(\kappa^1_4\)&
  \num{100}&
   \(\kappa^4_1\)&
  \num{2.0}\\
  (forward cycle) &%
   \(\kappa^4_3\)& \num{2.52e6}&
   \(\kappa^3_4=\frac{K_{\mathrm{eq}}\kappa^4_3\kappa^1_4\kappa^3_1}{\kappa^4_1 \kappa^1_3}\)&
  \num{49.3}\\
  \midrule Chemical transitions&%
   \(\kappa^3_2 = \left(\frac{\kappa^3_1}{\kappa^1_3}\right)^2\,\kappa^1_4\)&
  \num{6.4e-11}&
   \(\kappa^2_3=\kappa^4_1\)&
  \num{2.0}\\
  (backward cycle)&%
   \(\kappa^2_1=\kappa^4_3\)& \num{2.52e6}& \(\kappa^1_2=\kappa^3_4\)&
  \num{49.3}\\
  \midrule Mechanical load &%
   \(\chi^3_4=\chi^4_3=\chi^1_2=\chi^2_1\)&
  \num{0.15}&
   \(\chi^4_1=\chi^1_4=\chi^2_3=\chi^3_2\)&
  \num{0.25}\\
  \bottomrule
 \end{tabular}
 \caption{Numerical values of the parameters of the four-state model for kinesin.
  All first-order reaction rates \(\kappa\) are given in units of \si{\per\second} or, if attachment of $n$ chemicals is involved,
  \si{\per\second}\(\si{\micro}\si{\Molar}^{-n}\).
  Values correspond to the experimental data in Ref.~\cite{Carter.Cross2005} as stated in Ref.~\cite{Liepelt+Lipowsky2007}.
  The equilibrium constant of the ATP hydrolysis reaction is
   \(K_{\mathrm{eq}} = \SI{4.9e11}{\micro\Molar}\).
  The parameter \(\theta=\num{0.65}\) enters the mechanical factor of the transition rates.
 }
 \label{tab:parameters}
\end{table*}

The parametrization of the kinesin model on four states (\Fig{kinesin4state-cycles}) follows the steps in Ref.~\cite{Liepelt+Lipowsky2007} for the six-state model (\Fig{kinesin6state-cycles}).
Transition rates
\begin{align*}
 w^i_j \coloneqq \kappa^i_j\, C^i_j\, \Phi^i_j(f)
 \end{align*}
are obtained as first-order rate constants \(\kappa^i_j\), which are multiplied by concentration and force-dependent factors.
In accordance with first-order rate kinetics, the chemical factor reads
\begin{align*}
 C^i_j \coloneqq 
 \begin{cases}
  \prod_X[X] &
  \text{ if compound \(X\) is attached}.\\ 
  1 &
  \text{ else.}
 \end{cases}
\end{align*}

For chemical concentrations which are not too high, the non-dimensional chemical potential difference is given by
 \(\chempot=\ln \left(K_{\mathrm{eq}}\frac{[\mathrm{ATP}]}{[\mathrm{ADP}][\mathrm{P}]}\right)\),
where
 \(K_{\mathrm{eq}} = \SI{4.9e11}{\micro\Molar}\)
is the chemical equilibrium constant for the ATP hydrolysis reaction happening at kinesin's active sites.
Like in Ref.~\cite{Liepelt+Lipowsky2007} we fix  \([\text{ADP}]=[\text{P}]=\SI{1}{\micro\Molar}\) at physiological values and consequently vary the concentration of ATP as
\[
  [\mathrm{ATP}] = \frac{\mathrm{e}^{\chempot}}{49}\SI{e-10}{\micro\Molar}\,.
\]

The force dependent factors $\Phi^i_j$ depend on the non-dimensionalized pulling force
 \(f = L F /(k_\mathrm{B} T)\).
They have a different form for mechanical and chemical transitions:
\begin{align*}
 \Phi^i_j(f) \coloneqq
 \begin{cases}
  	 \exp{(-\theta f)}\,, & \text{if }(i\to j) = (1\to 3)\\
  \exp{( (1- \theta) f)}\,, &
  \text{if }(i\to j) = (3 \to 1),\\
  \frac{2}{1+\exp{[\chi^i_j f]}}\,, & \text{else,}
 \end{cases}
\end{align*}
where $\theta$ and \(\chi^i_j = \chi^j_i\) are additional experimental parameters.

For the six-state model the transitions of the forward cycle $\mathcal F = \zeta_1$ reflect experimental data. 
We briefly outline how we use the arguments of Ref.~\cite{Liepelt+Lipowsky2007} for the parametrization of the four-state model shown in \Fig{kinesin4state-cycles}.
First note that transitions associated to the edges $(1\leftrightarrow 3)$ and $(1\leftrightarrow4)$ are also present in the six-state model.
We thus use similar parameterizations.
Transition $(3\rightarrow 4)$ combines ADP release at the leading site with hydrolysis (and immediate P release) at the trailing one.
In the six-state model, the same numerical value of the mechanical parameter $\chi^i_j=0.15$ is used for both of these transitions.
We take this as a motivation for using $\chi^3_4=\chi^4_3=0.15$ in the four-state model.
Now the only undetermined parameters in the forward cycle are the first-order rate constants $\kappa^3_4$ and  $\kappa^4_3$.
The Hill--Schnakenberg condition $\circu{\sigma}{1}=-f+\chempot$ for vanishing mechanical driving $f=0$ yields one additional constraint which can be cast into the expression
\begin{align*}
 \frac{\kappa^3_4 \kappa^4_1\kappa^1_3}{\kappa^4_3\kappa^1_4\kappa^3_1} \stackrel{!}{=} K_{\mathrm{eq}}\,.
 \label{eq:balance}
\end{align*}
Finally, we take $\kappa^4_3$ as a fit parameter that we use to adjust our model to experimental results at the physiological parameter values:
we choose it such that the six-state and four-state model yield the same average velocity $\ssav{d}$ for $f=0$ and $[\text{ATP}]=[\text{ADP}]=[\text{P}] = \SI{1}{\micro\Molar}$.

The parameters of the remaining transitions are obtained by symmetry.
The exception is the first-order constant $\kappa^3_2$, associated with ATP release from the leading head.
Similar to Ref.~\cite{Liepelt+Lipowsky2007} we adjust it in order to account for the Hill--Schnakenberg conditions.
On the dissipative cycle  $\mathcal{D}=\zeta_2$ this constraint amounts to $\circu{\sigma}{2} = 2\chempot$ and yields  \(\kappa^3_2 = \left(\frac{\kappa^3_1}{\kappa^1_3}\right)^2\,\kappa^1_4\).

At this point, we have determined all the parameters of the four-state model while ensuring the physical and thermodynamic consistency with the original six-state model.
Fitting yields  $\kappa^4_3=2.52\times10^6$, which proves to be  a good choice globally, \cf{} \Fig{mismatch}.
All model parameters are summarized in Table~\ref{tab:parameters}.

\bibliography{current_cumulants}

\end{document}

%% file: current_cumulants_macros.tex
\newcommand{\reals}{\ensuremath{\mathbb{R}}}
\newcommand{\naturals}{\ensuremath{\mathbb{N}}}
\newcommand{\D}{\mathrm{d}}

\newcommand{\vertices}{\ensuremath{\mathcal{V}}}

\newcommand{\tredges}{\ensuremath{ {\mathcal{T}} }}  

\newcommand{\tray}{\ensuremath{\gamma}}
\newcommand{\obs}{\ensuremath{\varphi}}
\newcommand{\oobs}{\ensuremath{\psi}}
\newcommand{\bobs}[2]{\ensuremath{\obs_{(#1,#2)}}}

\newcommand{\chords}{\ensuremath{\mathcal{H}}}

\newcommand{\del}{\ensuremath{\partial}}

\newcommand{\abs}[1]{\ensuremath{\left|#1\right|}}

\DeclareMathOperator{\Cov}{Cov}

\newcommand{\circu}[2]{\ensuremath{\mathring{#1}_{#2}}}

\newcommand{\ie}{{\textit{i.\,e.\@{}}}}
\newcommand{\eg}{{\textit{e.\,g.\@{}}}}
\newcommand{\cf}{{\textit{cf.\@{}}}}

\newcommand{\tsub}[1]{\ensuremath{_{\mathrm{#1}}}} 

\newcommand{\Molar}{\textup{M}}		
\newcommand{\kb}{k\tsub{B}}

\renewcommand{\vec}{\bm}		

\newcommand{\chempot}{\Delta\mu}

\newcommand{\graph}{\mathcal G}

\newcommand{\trav}[1]{\left\langle #1 \right\rangle}
\newcommand{\ssav}[1]{c(#1)}
\newcommand{\tav}[1]{\overline{#1}_T}

\newcommand{\tmat}{\mathbb{W}}
\newcommand{\sfmat}[1]{\mathsf{#1}}

\theoremstyle{plain}
\theoremheaderfont{\normalfont\bfseries}
\theorembodyfont{\slshape}
\theoremseparator{}
\theoremsymbol{}

\theoremstyle{nonumberplain}

\theoremstyle{plain}
\theoremheaderfont{\normalfont\bfseries}
\theorembodyfont{\rmfamily}
\theoremseparator{}

\theoremheaderfont{\scshape}
\theorembodyfont{\upshape}
\theoremstyle{nonumberplain}
\theoremseparator{}
\theoremsymbol{\ensuremath{\square}}

\theoremstyle{plain}
\theoremheaderfont{\normalfont\bfseries}\theorembodyfont{\upshape}
\theoremseparator{}
\theoremsymbol{}

\newcommand{\revi}[1]{#1}

\newcommand\comment[1]{}

\newcommand{\eq}[1]{(\ref{eq:#1})}
\newcommand{\Eq}[1]{Eq.~\eq{#1}}

\newcommand{\fig}[1]{\ref{fig:#1}}
\newcommand{\Fig}[1]{Fig.~\fig{#1}}
\newcommand{\Figs}[1]{Figs.~\fig{#1}}

\newcommand{\Sec}[1]{Sec.~\ref{sec:#1}}

\include{revision}